\documentclass[conference]{IEEEtran}
\IEEEoverridecommandlockouts
\usepackage{cite}
\usepackage{amsmath,amssymb,amsfonts}
\usepackage{algorithmic}
\usepackage{graphicx}
\usepackage{textcomp}
\usepackage{xcolor}
\usepackage{tabularx} 
\usepackage{multirow}
\usepackage{graphicx,color}
\usepackage{textcomp}
\usepackage{hyperref}
\usepackage{cuted}
\usepackage{comment}
\def\BibTeX{{\rm B\kern-.05em{\sc i\kern-.025em b}\kern-.08em
    T\kern-.1667em\lower.7ex\hbox{E}\kern-.125emX}}
\begin{document}

\title{Performance Analysis of MDMA-Based Cooperative MRC Networks with Relays in Dissimilar Rayleigh Fading Channels\\
\thanks{This work was supported by Natural Science Foundation of China (Project
	Number: U22B2003 and U2001208). (Corresponding author: Chen Dong).\\
	Lei Teng, Wannian An, Chen Dong, Xiaoqi Qin and Xiaodong Xu are with the School of Information and Communication Engineering
	Beijing University of Posts and Telecommunications.}
}

\author{\relax Lei Teng\IEEEauthorrefmark{2}, Wannian An\IEEEauthorrefmark{2}, Chen Dong\IEEEauthorrefmark{2},   Xiaoqi Qin\IEEEauthorrefmark{2}, Xiaodong Xu\IEEEauthorrefmark{2} \\
	\IEEEauthorblockA{\IEEEauthorrefmark{2}State Key Laboratory of Networking and Switching Technology,\\ Beijing University of Posts and Telecommunications, Beijing, 100876, China. \\ Email: tenglei@bupt.edu.cn, anwannian2021@bupt.edu.cn, dongchen@bupt.edu.cn,\\ xiaoqiqin@bupt.edu.cn, xuxiaodong@bupt.edu.cn}
	 }

\maketitle

\begin{abstract}
Multiple access technology is a key technology in various generations of wireless communication systems. As a potential multiple access technology for the next generation wireless communication systems, model division multiple access (MDMA) technology improves spectrum efficiency and feasibility regions. This implies that the MDMA scheme can achieve greater performance gains compared to traditional schemes. Relay-assisted cooperative networks, as a infrastructure of wireless communication, can effectively utilize resources and improve performance when MDMA is applied. In this paper, a communication relay cooperative network based on MDMA in dissimilar rayleigh fading channels is proposed, which consists of two source nodes, any number of decode-and-forward (DF) relay nodes, and one destination node,  as well as using the maximal ratio combining (MRC) at the destination to combine the signals received from the source and relays. By applying the state transition matrix (STM) and moment generating function (MGF), closed-form analytical solutions for outage probability and resource utilization efficiency are derived. Theoretical and simulation results are conducted to verify the validity of the  theoretical analysis.
\end{abstract}


\section{Introduction}
Semantic communication is considered a potential paradigm for next-generation communication systems\cite{b1}, which involves the selective extraction, compression, transmission of features from the original signals, and the utilization of semantic-level information for communication purposes.

 The goal of the text-based semantic communication system \cite{b2}, namely the deep learning based semantic communication system (DeepSC), is to recover the meaning of sentences in traditional communication systems rather than bit or symbol errors, with the aim of maximizing system capacity and minimizing semantic errors. It exhibits significant semantic transmission advantages under low signal-to-noise ratio (SNR) conditions compared to traditional communication systems. According to \cite{b3},   the layer-based semantic communication system for image (LSCI) model serves as an intelligent carrier, and semantic transmission is essentially the propagation of artificial intelligence models. Therefore, the semantic slice model (SeSM) is designed to achieve semantic intelligent propagation. In \cite{b4}, nonlinear transform source-channel coding (NTSCC) is applied to map image and video sources into a nonlinear latent space for more efficient semantic extraction and transmission.

 The aforementioned semantic communication systems are designed for point-to-point communication, while multi-user communication systems have always been a focus of research \cite{b5}. From the first generation (1G) to the fifth generation (5G) of communication system development, each generation has featured representative and groundbreaking multiple access (MA) technologies \cite{b6}. A multi-user semantic communication system based on Non-Orthogonal Multiple Access (NOMA) is proposed in \cite{b7}, which supports semantic transmission for multiple users with different source information modalities. In order to harness the performance potential of semantic information, a novel multiple access technology based on semantic domain resources is proposed in \cite{b8}, known as Model Division Multiple Access (MDMA). It explores the shared and personalized information from a high-dimensional semantic space. The shared semantic information is transmitted within the same time-frequency resources, while personalized semantic information is transmitted separately. Compared to other MA technologies, MDMA's gain primarily stems from the reuse of shared information among different users in the model information space.

 Furthermore, relay-assisted cooperative networks are an important research area in communication systems. In \cite{b9}, an exact closed-form expression for the outage probability of DF relay in a communication network with m relays is derived in different Rayleigh fading channels. In \cite{b10}, a cooperative communication network based on energy harvesting (EH) and DF relaying is considered. A STM is proposed to obtain the probability distribution of the candidate broadcast node set. The combination of relay-assisted cooperative networks and multiple access technology is also a research focus. In \cite{b11}, a time division multiple access (TDMA)-based cooperative medium access control protocol for wireless multi-hop relaying networks is proposed and analyzed. In \cite{b12}, the impact of relay selection (RS) on the performance of cooperative NOMA is studied. In addition, diversity gain techniques are of significant importance for relay-assisted cooperative networks. In \cite{b13}, the performance of NOMA in a collaborative system is studied when a source node communicates with two EH user devices using MRC technique with multiple antenna hybrid AF/DF relay nodes. In \cite{b14}, a cooperative communication network based on two energy harvesting DF relays is proposed, where the diversity gain is obtained at the destination node using MRC technique, and the probability density function of the total SNR collected at the destination node under different transmitter-receiver node states is calculated using STM.

 Based on the above, it can be observed that the combination of multiple access technologies and cooperative networks for next-generation communication systems has received less attention. This paper aims to propose a communication cooperative network utilizing MDMA. The main contributions of this paper are as follows:

\begin{enumerate}
	\item A DF relays-assisted MRC network based on MDMA, which consists of two source nodes, an arbitrary number of DF relay nodes, and one destination node, is proposed.
	\item By utilizing MGF and STM, closed-form theoretical expressions for the outage probability, resource utilization efficiency, and the average number of time slots required for each data transmission are derived in this network.
	\item Through numerous simulation results, it is demonstrated that simulation results match with corresponding theoretical results.
	
\end{enumerate}

$Notations:$

\begin{table}[h!]
	\caption{Notations table}
	\label{tab:1} 
	\begin{tabular}{|c|p{5.8cm}<{\centering}|}
		\hline
		Notation & Meaning \\
		\hline
		$A\sim\mathcal{CN}(0, \theta)$& the random variable $A$ follows the complex Gaussian distribution with mean 0 and variance $\theta$\\
		\hline
		Boldface capital $\textbf{T}$ & matrices or vectors\\
		\hline
		\multirow{3}{*}{$|$}&conditional probability. For example, case A,B$|$C means event A and B occur under the condition of C.\\
		\hline
		\multirow{2}{*}{conv$(A,B)$}&discrete convolution $y(j)=\mathrm{conv}(A,B)=\sum_{i=-\infty}^{+\infty}A(i)B(j-i)$\\
		\hline
		$\lceil x \rceil$&$x$ upwards to the nearest integer\\
		\hline
	\end{tabular}
\end{table}
In TABLE \ref{tab:1}, all notations are showed. \\

The remaining paper is developed as follows:
Section II interprets the system model. In section III, the expressions for the outage probability, resource utilization efficiency and the average number of time slots required for each image transmission are determined. In section IV, simulation performance results are displayed. Section V is where the conclusion is provided.

\section{System Model}

\begin{figure}[h]
	
	\centerline{\includegraphics[width=3.5in]{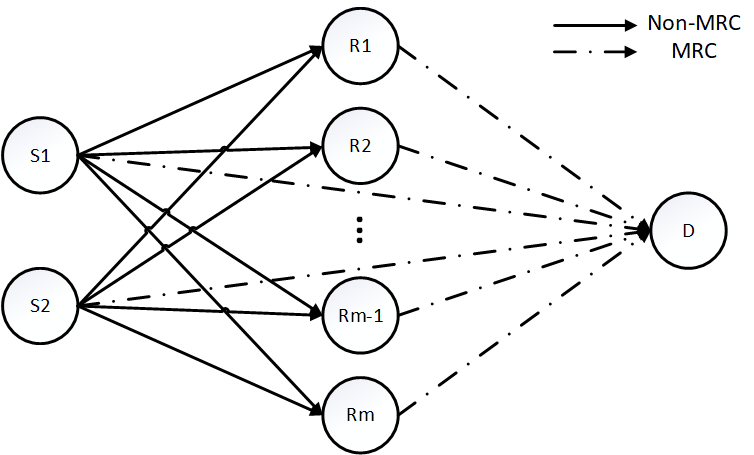}}
	\caption{System model\label{system}}
\end{figure}

Consider the wireless network depicted in Fig \ref{system}, where the source nodes S1 and S2 employ a set of m relay nodes $c = \{R1, ..., Rm\}$. Let $d_{ab}$ denote the distance between nodes with a and b. The mutually independent complex channel gains between the nodes in the $j$-th time slot are modeled as $h_{S1D}(j)\sim\mathcal{CN}(0, d_{S1D}^{-\alpha})$, $h_{S2D}(j)\sim\mathcal{CN}(0, d_{S2D}^{-\alpha})$, $h_{S1Ri}(j)\sim\mathcal{CN}(0, d_{S1Ri}^{-\alpha})$, $h_{S2Ri}(j)\sim\mathcal{CN}(0, d_{S2Ri}^{-\alpha})$, $h_{RiD}(j)\sim\mathcal{CN}(0, d_{RiD}^{-\alpha})$, where $i = 1, ..., m$ and the term $\alpha$ represents the path-loss exponent.

 The source S1 broadcasts unit-energy symbols $x_{S1}(i)$ to relay Ri and destination D at rate $R_0$ with the constant power $P_S$. The received signals $y_{S1Ri}(j)$, and $y_{SD}(j)$ at relay Ri and destination D in the $j$-th time slot are given by
 \begin{equation}
 	y_{S1Ri}(j) = \sqrt{P_{S}}h_{S1Ri}(j)x_S(j) + n_{S1Ri}(j),
 \end{equation}
 \begin{equation}
 	y_{S1D}(j) = \sqrt{P_{S}}h_{S1D}(j)x_S(j) + n_{S1D}(j),
 \end{equation}
 where, $n_{S1Ri}(j)$ and $n_{S1D}(j)\sim\mathcal{CN}(0, N_0)$ represent the received additive white Gaussian noise (AWGN) at Ri and D, respectively. Thus, the received signal-to-noise ratios (SNRs) $\gamma_{S1Ri}(j)$ and $\gamma_{S1D}(j)$ at Ri and D in the $j$-th time slot would be given as follows
 \begin{equation}
 	\begin{split} 
 		 \gamma_{S1Ri}(j) = \frac{P_S|h_{S1Ri}(j)|^2}{N_0}
 		 \mbox{ and }  \gamma_{S1D}(j) = \frac{P_S|h_{S1D}(j)|^2}{N_0}.   
 	\end{split}
 \end{equation}
 \begin{figure}[h]
 	
 	\centerline{\includegraphics[width=3.5in]{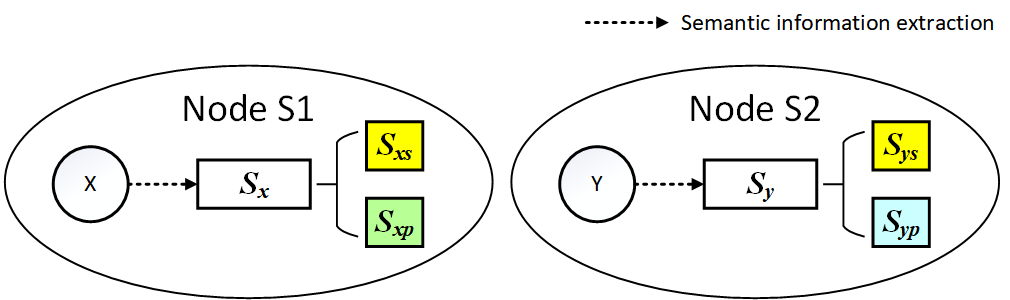}}
 	\caption{The shared information $S_{xs}$ and $S_{ys}$ and the personalized information $S_{xp}$ and $S_{yp}$ extraction model\label{MDMA}}
 \end{figure}
 
 Similarly, 
  \begin{equation}
 	y_{S2Ri}(j) = \sqrt{P_{S}}h_{S2Ri}(j)x_S(j) + n_{S2Ri}(j),
 \end{equation}
 \begin{equation}
 	y_{S2D}(j) = \sqrt{P_{S}}h_{S2D}(j)x_S(j) + n_{S2D}(j),
 \end{equation}
  \begin{equation}
 	y_{RiD}(j) = \sqrt{P_{S}}h_{RiD}(j)x_S(j) + n_{RiD}(j),
 \end{equation}
  \begin{equation}
 	\begin{split} 
 		&\gamma_{S2Ri}(j) = \frac{P_S|h_{S2Ri}(j)|^2}{N_0},
 		\gamma_{S2D}(j) = \frac{P_S|h_{S2D}(j)|^2}{N_0} 
 		\\ &\qquad\qquad\mbox{ and } 
 		  \gamma_{RiD}(j) = \frac{P_S|h_{RiD}(j)|^2}{N_0}.
 	\end{split}
 \end{equation}
 Let 
 \begin{equation}
 	\Gamma_{th} = 2^{R_0}-1
 \end{equation}
 be the network SNR threshold, where $R_0$ represents the target data transmitting rate. When the SNR is greater than $\Gamma_{th}$, the receiver is assumed to successfully decode the information.
 
 Furthermore, we assume that the receivers at the relays and the destination have accurate channel state information, enabling MRC. However, there is no transmitter channel state information available at the source or relays. In addition, in Fig.\ref{MDMA}, S1 and S2 extract their source semantic information $S_x$ and $S_y$, respectively, as follows: 
\begin{equation}
	S_x = \phi_1(X), S_y = \phi_2(Y),
\end{equation}
 where $S_x$ and $S_y$ represent the semantic information extracted from X and Y, respectively. We define the decoding set $C$ as the set of relays in $c$ that have the ability to successfully decode the source messages. As shown in \cite{b7}, the shared information $S_{xs}$ and $S_{ys}$ and the personalized information $S_{xp}$ and $S_{yp}$ can be extracted. With the help of MDMA, D only needs to successfully receive  $S_{xs}$, $S_{xp}$ and $S_{yp}$ to successfully reconstruct X and Y. This is because $S_{xs}$ and $S_{ys}$ have a high degree of similarity. The ratio of the number of shared information bits $B_s$ to the total number of information bits $B_t$ is defined as $\eta$, and correspondingly, $1-\eta$ represents the ratio of the number of personalized information bits $B_p$ to the total number of information bits $B_t$. As a result, the number of time slots required for transmitting shared information $\beta_s$ and personalized information $\beta_p$ can be defined as follows:
\begin{equation}\label{eq1}
	\beta_s = \lceil \frac{B_s}{R_0}\rceil = \lceil\frac{\eta B_t}{R_0}\rceil,\beta_p = \lceil \frac{B_p}{R_0}\rceil = \lceil\frac{(1-\eta )B_t}{R_0}\rceil,
\end{equation}
To meet the multiple access requirements of S1 and S2, MDMA is applied in the system. The information transmission in MDMA is divided into two phases:

Phase I:

\qquad step1: S1 broadcasts the shared semantic information $S_{xs}$ to D. If $\gamma_{S1D}$ is greater than the threshold $\Gamma_{th}$, D successfully decodes the information and broadcast positive acknowledgement (ACK). Otherwise, step2 proceeds in the next time slot. If D has already successfully received $\beta_s$ times the shared information, Phase II proceeds in the next time slot. Otherwise, repeat Phase I in the next time slot.

\qquad step2: If $C$ is empty, it is recorded as a failure and repeat step1 in the current time slot. Otherwise, the relays in $C$ simultaneously transmit it to D. If the total SNR exceeds the threshold through MRC, D is considered to successfully decode the information and broadcast ACK. If D has already successfully received $\beta_s$ times the shared information, Phase II begins. Otherwise, repeat step1 in the next time slot.

Phase II: S1 and S2 use TDMA to transmit personalized semantic information ($S_{xp}$ and $S_{yp}$) to D. First, S1 transmits its information $S_{xp}$, followed by S2's information $S_{yp}$. The personalized transmission process is consistent with Phase I but $\beta_p$ times. Once $S_{xp}$ and $S_{yp}$ are both transmitted, S1 and S2 send new information.

\section{Performance of DF Relaying in Rayleigh Fading Channels}
	To obtain theoretical performance expressions for the system, we can calculate the performance for different steps in MDMA separately. For Phase I step1 ($pIS1s1$ for short), the outage probability $OP_{pIS1s1}$ can be expressed as follows:
	\begin{equation}
		OP_{pIS1s1} = \mathrm{Pr}\{\gamma_{S1D} < \Gamma_{th}\} = 1-e^{-d_{S1D}^{\alpha}\Gamma_{th}/SNR}.
	\end{equation}
	For Phase I step2, the combined SNR $\gamma_{combined}$ can be expressed as follows:
		\begin{equation}
		\gamma_{combined1} =\gamma_{S1D}+\sum_{Ri\in C}\gamma_{RiD}
	\end{equation}
	 Through the Theorem of Total Probability, the outage probability $OP_{pIS1s2}$ can be given by
	\begin{equation}\label{eq13}
		OP_{pIS1s2} = \frac{\sum_{C}^{C \neq \emptyset} \mathrm{Pr}\{\gamma_{combined} < \Gamma_{th} | \gamma_{S1D} < \Gamma_{th} \}\mathrm{Pr}\{C\}}{1-\mathrm{Pr}\{C = \emptyset\}}.
	\end{equation}
	Let path 0 represent the direct link from S1 to D, and path i represent the cascaded link from S1 to Ri to D, where i = 1, ..., m. Let random variable $y_{S1,i}$ denote the square gain on the i-th cascaded link.
	The random variable $y_{S1,i}$ takes into account both the fading on the source-to-i-th relay link and the fading on the i-th relay-to-destination link. Then, $y_{S1,i}$ has a probability density function (PDF)
	\begin{equation}
		\begin{split}
		f_{y_{S1,i}}(x)=
		& f_{y_{S1,i}|\mathrm{negative}}\mathrm{Pr}\{\mathrm{negative}\}\\
		& +f_{y_{S1,i}|\mathrm{positive}}\mathrm{Pr}\{\mathrm{positive}\}.
	\end{split}
	\end{equation}
	If the i-th path is negative, the conditional PDF $f_{y_{S1,i}|\mathrm{negative}} = \delta(0)$, where $\delta(0)$ represents the Dirac Delta function, which is a pulse function at zero. The probability of this event occurring is denoted as $A_i$:
	\begin{equation}
		A_i = \mathrm{Pr}\{\gamma_{S1Ri} < \Gamma_{th}\} = 1-e^{-d_{S1Ri}^{\alpha}\Gamma_{th}/SNR}.
	\end{equation}
	Obviously, the probability that the i-th path is positive is $1-A_i$ and 
	\begin{equation}
		f_{y_{S1,i}|\mathrm{positive}}(x)= \frac{d_{RiD}^{\alpha}}{SNR}e^{-\frac{d_{RiD}^{\alpha}}{SNR}x}.
	\end{equation}
	Therefore, 
		\begin{equation}
		\begin{split}
			f_{y_{S1,i}}(x)= \delta(0)A_i+\frac{d_{RiD}^{\alpha}}{SNR}e^{-\frac{d_{RiD}^{\alpha}}{SNR}x}(1-A_i)\\
			i=1, 2,..., m.
		\end{split}
	\end{equation}
	\begin{figure*}[!t]\centering
		\begin{equation}\label{F}\tag{20}
			\begin{split}
				F_{sum}(\gamma) =& \sum_{j=1}^m \frac{(1-A_j)\prod_{k=1}^m A_k}{A_j}[1-e^{-\frac{d_{RjD}^{\alpha}}{SNR}\gamma}]+\sum_{j=1}^m\sum_{r=j+1}^m \frac{(1-A_j)(1-A_r)\prod_{k=1}^m A_k}{A_j A_r}[\theta_{j,k}(1-e^{-\frac{d_{RjD}^{\alpha}}{SNR}\gamma})\\
				&+\theta_{k,j}(1-e^{-\frac{d_{RkD}^{\alpha}}{SNR}\gamma})]
				+...+(1-A_1)(1-A_2)(1-A_3)...(1-A_m)\big[\theta_{1,2}\theta_{1,3}...\theta_{1,m}(1-e^{-\frac{d_{R1D}^{\alpha}}{SNR}\gamma})\\
				&+\theta_{2,1}\theta_{2,3}...\theta_{2,m}(1-e^{-\frac{d_{R2D}^{\alpha}}{SNR}\gamma})+...+\theta_{m,1}\theta_{m,2}...\theta_{m,m-1}(1-e^{-\frac{d_{RmD}^{\alpha}}{SNR}\gamma})\big],
			\end{split}
		\end{equation}
		\hrulefill
	\end{figure*}
	To obtain Eq.(\ref{eq13}), the CDF of $\sum_{C}^{C \neq \emptyset} \sum_{Ri\in C}\gamma_{RiD}\mathrm{Pr}\{C\}$ can be gotten firstly. The PDF can be determined through the moment generating functions, and then obtain the CDF from the PDF by integration. The MGF of $i$-th path is
	\begin{equation}
		M_i(s) = A_i+(1-A_i)\frac{\frac{d_{RiD}^{\alpha}}{SNR}}{s+\frac{d_{RiD}^{\alpha}}{SNR}}.
	\end{equation}
	Due to the mutual independence of $y_{S1,i}$, the MGF of $\sum_{C}^{C \neq \emptyset} \sum_{Ri\in C}\gamma_{RiD}\mathrm{Pr}\{C\}$ can be expressed as follows:
		\begin{equation}
		M_{sum}(s) = \prod_{i=1}^m M_i(s).
	\end{equation}
		
	After some simplification and applying the Laplace inverse transform, the CDF can be obtained as Eq.(\ref{F}),
	where 
	\begin{equation}\setcounter{equation}{21}
		\begin{split}
			\theta_{x,y}=\frac{d_{RyD}^{\alpha}}{d_{RyD}^{\alpha}-d_{RxD}^{\alpha}}.
		\end{split}
	\end{equation}
		
	Let $p_{\gamma_{overall1}}(j)$ be the probability which are random variable $\sum_{C}^{C \neq \emptyset} \sum_{Ri\in C}\gamma_{RiD}$ of value $(\frac{j-1}{\mathcal{N}}\Gamma_{th}\leq \sum_{C}^{C \neq \emptyset} \sum_{Ri\in C}\gamma_{RiD}<\frac{j}{\mathcal{N}}\Gamma_{th})$. Hence,
	\begin{equation}
		\begin{split}
			p_{\gamma_{overall1}}(j)=F_{sum}(\frac{j}{\mathcal{N}}\Gamma_{th})-F_{sum}(\frac{j-1}{\mathcal{N}}\Gamma_{th})\\
			j=1,2,...,\mathcal{N}.
		\end{split}
	\end{equation}

where $\mathcal{N}$ represents the granularity of differentiation. Similarly, $p_{\gamma_{S1D}}(j)$ is the probability which are random variable $\gamma_{S1D}$ of value $\{(\frac{j-1}{\mathcal{N}}\Gamma_{th}\leq \gamma_{S1D}<\frac{j}{\mathcal{N}}\Gamma_{th})|\gamma_{S1D} < \Gamma_{th}\}$,
\begin{equation}
	\begin{split}
		p_{\gamma_{S1D}}(j)=\frac{e^{-d_{S1D}^{\alpha}\frac{j-1}{\mathcal{N}}\Gamma_{th}/SNR}-e^{-d_{S1D}^{\alpha}\frac{j}{\mathcal{N}}\Gamma_{th}/SNR}}{1-e^{-d_{S1D}^{\alpha}\Gamma_{th}/SNR}}\\
		j=1,2,...,\mathcal{N}.
	\end{split}
\end{equation}
Let $p_{\gamma_{combined1}}(j)$ be the probability which are random variable $\sum_{C}^{C \neq \emptyset}\{\gamma_{combined} < \Gamma_{th} | \gamma_{S1D} < \Gamma_{th} \}$ of value $(\frac{j-1}{\mathcal{N}}\Gamma_{th}\leq \sum_{C}^{C \neq \emptyset}\{\gamma_{combined} < \Gamma_{th} | \gamma_{S1D} < \Gamma_{th} \}<\frac{j}{\mathcal{N}}\Gamma_{th})$. Obviously, 
\begin{equation}
	\begin{split}
		p_{\gamma_{combined1}}= \mathrm{conv}(p_{\gamma_{overall1}},p_{\gamma_{S1D}}).
	\end{split}
\end{equation}
and Eq.(\ref{eq13})  can be written as follows
\begin{equation}
	\begin{split}
		OP_{pIS1s2}= \frac{\sum_{j=1}^{\mathcal{N}}p_{\gamma_{combined1}}(j)}{1-\prod_{i=1}^{m}A_i}.
	\end{split}
\end{equation}

According to the system model, the transmission process of Phase II S1 step 1 and Phase II S1 step 2 is similar to that of Phase I step 1 and Phase I step 2, with the exception of varying sizes of information. Therefore, the outage probability $OP_{pIIS1s1}$ and $OP_{pIIS1s2}$ can be given by
\begin{equation}
	\begin{split}
		OP_{pIIS1s1}= OP_{pIS1s1}, OP_{pIIS1s2}=OP_{pIS1s2}.
	\end{split}
\end{equation}

Regarding Phase II S2 step 1 and Phase II S2 step 2, similar to the calculation of Phase I step 1 and Phase I step 2 mentioned earlier, obtaining $OP_{pIIS2s1}$ and $OP_{pIIS2s2}$ only requires replacing S1 with S2 in the calculation formulas.

Thus, the expressions of the outage probability for each phase and each step are obtained. The final expression for the outage probability $OP$ is as follows:
\begin{equation}\label{eq27}
	\begin{split}
		OP=& OP_{pIS1s1}\sum_{i=1}^{\beta_s}p_{\mathrm{pIS1s1},i}+OP_{pIS1s2}\sum_{i=1}^{\beta_s}p_{\mathrm{pIS1s2},i}\\
		&+OP_{pIIS1s1}\sum_{i=1}^{\beta_p}p_{\mathrm{pIIS1s1},i}+OP_{pIIS1s2}\sum_{i=1}^{\beta_p}p_{\mathrm{pIIS1s2},i}\\
		&+OP_{pIS2s1}\sum_{i=1}^{\beta_p}p_{\mathrm{pIIS2s1},i}+OP_{pIS2s2}\sum_{i=1}^{\beta_p}p_{\mathrm{pIIS2s2},i},
	\end{split}
\end{equation}
where $p_{\mathrm{pIS1s1},i}$ represents the probability of phase I $i$-th S1 transmitting step1 and so on. Through STM, these probabilities can be obtained.

Let $p$ be the probability distribution 
\begin{equation}
	\begin{split}
		p(1)=&[p_{\mathrm{pIS1s1},1}, p_{\mathrm{pIS1s2},1},p_{\mathrm{pIS1s1},2}, p_{\mathrm{pIS1s2},2},...,p_{\mathrm{pIS1s1},\beta_s},\\
		&p_{\mathrm{pIS1s2},\beta_s},p_{\mathrm{pIIS1s1},1},p_{\mathrm{pIIS1s2},1},...,p_{\mathrm{pIIS1s1},\beta_p},p_{\mathrm{pIIS1s2},\beta_p},\\
		&p_{\mathrm{pIIS2s1},1},p_{\mathrm{pIIS2s2},1},...,p_{\mathrm{pIIS2s1},\beta_p},p_{\mathrm{pIIS2s2},\beta_p}].
	\end{split}
\end{equation}
Using STM,
\begin{equation}
	\begin{split}
		p(i+1)=p(i)\textbf{T},
	\end{split}
\end{equation}

where 
		\begin{strip}
\begin{equation}\small
	\begin{split}
		\textbf{T}=\begin{bmatrix}
			p_{(\mathrm{pIS1s1},1)-(\mathrm{pIS1s1},1)}      & p_{(\mathrm{pIS1s1},1)-(\mathrm{pIS1s2},1)}  & p_{(\mathrm{pIS1s1},1)-(\mathrm{pIS1s1},2)}  & ...  & p_{(\mathrm{pIS1s1},1)-(\mathrm{pIIS2s1},\beta_p)}   & p_{(\mathrm{pIS1s1},1)-(\mathrm{pIIS2s2},\beta_p)}  \\
		p_{(\mathrm{pIS1s2},1)-(\mathrm{pIS1s1},1)}      & p_{(\mathrm{pIS1s2},1)-(\mathrm{pIS1s2},1)}  & p_{(\mathrm{pIS1s2},1)-(\mathrm{pIS1s1},2)}  & ...  & p_{(\mathrm{pIS1s2},1)-(\mathrm{pIIS2s1},\beta_p)}   & p_{(\mathrm{pIS1s2},1)-(\mathrm{pIIS2s2},\beta_p)}  \\
			p_{(\mathrm{pIS1s1},2)-(\mathrm{pIS1s1},1)}      & p_{(\mathrm{pIS1s1},2)-(\mathrm{pIS1s2},1)}  & p_{(\mathrm{pIS1s1},2)-(\mathrm{pIS1s1},2)}  & ...  & p_{(\mathrm{pIS1s1},2)-(\mathrm{pIIS2s1},\beta_p)}   & p_{(\mathrm{pIS1s1},2)-(\mathrm{pIIS2s2},\beta_p)}  \\
			...      & ...  & ...  & ...  & ...   & ...  \\
			p_{(\mathrm{pIIS2s1},\beta_p)-(\mathrm{pIS1s1},1)}      & p_{(\mathrm{pIIS2s1},\beta_p)-(\mathrm{pIS1s2},1)}  & p_{(\mathrm{pIIS2s1},\beta_p)-(\mathrm{pIS1s1},2)}  & ...  & p_{(\mathrm{pIIS2s1},\beta_p)-(\mathrm{pIIS2s1},\beta_p)}   & p_{(\mathrm{pIIS2s1},\beta_p)-(\mathrm{pIIS2s2},\beta_p)}  \\
			p_{(\mathrm{pIIS2s2},\beta_p)-(\mathrm{pIS1s1},1)}      & p_{(\mathrm{pIIS2s2},\beta_p)-(\mathrm{pIS1s2},1)}  & p_{(\mathrm{pIIS2s2},\beta_p)-(\mathrm{pIS1s1},2)}  & ...  & p_{(\mathrm{pIIS2s2},\beta_p)-(\mathrm{pIIS2s1},\beta_p)}   & p_{(\mathrm{pIIS2s2},\beta_p)-(\mathrm{pIIS2s2},\beta_p)}  \\
		\end{bmatrix},
	\end{split}
\end{equation}
where $p_{(\mathrm{pIS1s1},1)-(\mathrm{pIS1s2},1)}$ represents the probability of a state transition from $(\mathrm{pIS1s1},1)$ to $(\mathrm{pIS1s2},1)$ and so on.
		\end{strip}
According to the system model, we have
\begin{equation}
	\begin{split}
		&p_{(\mathrm{pIS1s1},j)-(\mathrm{pIS1s1},j)}=OP_{pIS1s1}\prod_{i=1}^{m}A_i\\
		&p_{(\mathrm{pIS1s1},j)-(\mathrm{pIS1s2},j)}=OP_{pIS1s1}(1-\prod_{i=1}^{m}A_i)\\& \qquad \qquad\qquad \qquad\qquad \qquad\qquad j=1,2,...,\beta_s,
	\end{split}
\end{equation}
\begin{equation}
	\begin{split}
		p_{(\mathrm{pIS1s2},j)-(\mathrm{pIS1s1},j)}&=OP_{pIS1s2}\\
		p_{(\mathrm{pIS1s2},j)-(\mathrm{pIS1s1},j+1)}&=1-OP_{pIS1s2}\\
		p_{(\mathrm{pIS1s1},j)-(\mathrm{pIS1s1},j+1)}&=1-OP_{pIS1s1}
		\\&    \qquad j=1,2,...,\beta_s-1,
	\end{split}
\end{equation}
\begin{equation}
	\begin{split}
		p_{(\mathrm{pIS1s1},\beta_s)-(\mathrm{pIIS1s1},1)}=1-OP_{pIS1s1},
	\end{split}
\end{equation}

\begin{equation}
	\begin{split}
		&p_{(\mathrm{pIIS1s1},j)-(\mathrm{pIIS1s1},j)}=OP_{pIIS1s1}\prod_{i=1}^{m}A_i\\
		&p_{(\mathrm{pIIS1s1},j)-(\mathrm{pIIS1s2},j)}=OP_{pIIS1s1}(1-\prod_{i=1}^{m}A_i)\\& \qquad \qquad\qquad \qquad\qquad \qquad\qquad j=1,2,...,\beta_p,
	\end{split}
\end{equation}
\begin{equation}
	\begin{split}
		p_{(\mathrm{pIIS1s2},j)-(\mathrm{pIIS1s1},j)}&=OP_{pIIS1s2}\\
		p_{(\mathrm{pIIS1s2},j)-(\mathrm{pIIS1s1},j+1)}&=1-OP_{pIIS1s2}\\
		p_{(\mathrm{pIIS1s1},j)-(\mathrm{pIIS1s1},j+1)}&=1-OP_{pIIS1s1}
		\\&    \qquad j=1,2,...,\beta_p-1,
	\end{split}
\end{equation}
\begin{equation}
	\begin{split}
		p_{(\mathrm{pIIS1s1},\beta_p)-(\mathrm{pIIS1s1},1)}=1-OP_{pIIS1s1},
	\end{split}
\end{equation}

\begin{equation}
	\begin{split}
		&p_{(\mathrm{pIIS2s1},j)-(\mathrm{pIIS2s1},j)}=OP_{pIIS2s1}\prod_{i=1}^{m}\hat{A}_i\\
		&p_{(\mathrm{pIIS2s1},j)-(\mathrm{pIIS2s2},j)}=OP_{pIIS2s1}(1-\prod_{i=1}^{m}\hat{A}_i)\\& \qquad \qquad\qquad \qquad\qquad \qquad\qquad j=1,2,...,\beta_p,
	\end{split}
\end{equation}
\begin{equation}
	\begin{split}
		p_{(\mathrm{pIIS2s2},j)-(\mathrm{pIIS2s1},j)}&=OP_{pIIS2s2}\\
		p_{(\mathrm{pIIS2s2},j)-(\mathrm{pIIS2s1},j+1)}&=1-OP_{pIIS2s2}\\
		p_{(\mathrm{pIIS2s1},j)-(\mathrm{pIIS2s1},j+1)}&=1-OP_{pIIS2s1}
		\\&    \qquad j=1,2,...,\beta_p-1,
	\end{split}
\end{equation}
\begin{equation}
	\begin{split}
		p_{(\mathrm{pIIS2s1},\beta_p)-(\mathrm{pIS1s1},1)}=1-OP_{pIIS2s1},
	\end{split}
\end{equation}
\begin{equation}
	\begin{split}
	\hat{A}_i = \mathrm{Pr}\{\gamma_{S2Ri} < \Gamma_{th}\} = 1-e^{-d_{S2Ri}^{\alpha}\Gamma_{th}/SNR}\\i=1,2,...,m,
	\end{split}
\end{equation}
and the remaining elements In $\textbf{T}$ that have not been mentioned are all assumed to be zero.

Based on the algorithm 1 described in \cite{b10} and utilizing $\textbf{T}$ and $p(1)$, we can obtain the final probability distribution of the states. By substituting this distribution into Eq.(\ref{eq27}), we can derive the closed-form analytical expression for the outage probability.

Furthermore, the time slot cost for each data $T_c$ is a performance metric worthy of attention, which can be calculated as follows:
\begin{equation}\footnotesize
	\begin{split}
		T_c  & = \lim_{N \to \infty}\sum_{i=1}^N OP^{i-1}(1-OP)i\\
		&=\lim_{N \to \infty} (1-OP) \frac{1}{1-OP}[1+\frac{(OP)(1-(OP)^{N-1})}{1-OP}\\
		&-N(OP)^{N}]\\
		&=\frac{1}{1-OP}.
	\end{split}
\end{equation}

In addition, the resource utilization efficiency of the system $\varphi$ can be defined as:
\begin{equation}
	\begin{split}
		\varphi  & = \frac{2}{T_c(\beta_s+2\beta_p)BW},
	\end{split}
\end{equation}
where $B$ represents the size of the utilized bandwidth, while W represents the size of the utilized power.

\section{Simulation Performance Results}
In this section, simulation results are presented to demonstrate the validity of the derived theoretical analytical expressions. Additionally, performance comparisons are made between the cooperative network employing MDMA and the cooperative networks employing NOMA, FDMA, or TDMA. For all simulations, the following system parameters are taken into account unless otherwise specified. Suppose S1, S2, Ri (i=1,2,...,8) and D are all located on a two-dimensional plane, and their position coordinates are $(20,20), (0,20), (50,50-100(i-0.5)/8+5)$, and $(100,0)$, respectively. Path-loss exponent $\alpha = 3$. The target data transmitting rate $R_0=1$ bit/s/Hz. The total number of one image information bits $B_t=10$ bits. The channel noise variance $N_0=-50$dBm.  Furthermore, in all figures, blue markers represent simulation values, and red lines indicate the STM-based theoretical values, while the performance of other MA systems is based on simulation results.

\begin{figure}[h!]
	\centerline{\includegraphics[width=3.5in]{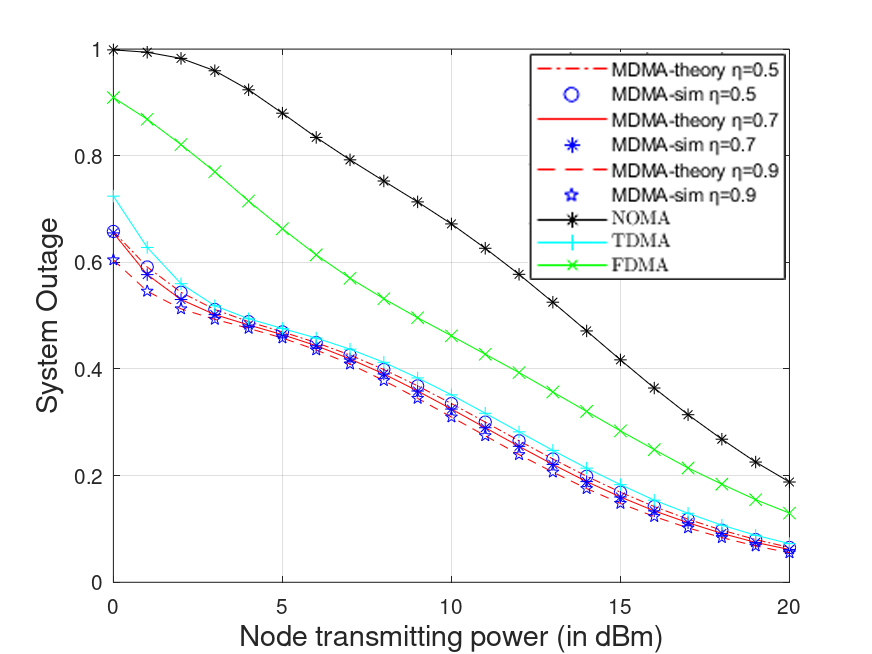}}
	\caption{Outage probability of system vs. node transmitting power.}
	\label{fig1}
\end{figure}
In Fig.\ref{fig1}, it can be observed that the Outage probability of the cooperative network employing MDMA is the lowest among all the compared systems. Moreover, as the parameter $\eta$ approaches 1, the outage probability decreases further. This can be attributed to the fact that S1 is closer to the relay and destination nodes, resulting in a higher proportion of transmissions utilizing S1 in the MDMA-based network. As $\eta$ increases, the usage of S1 for transmission becomes more dominant.
Additionally, the outage probability of the TDMA-based network is similar to that of the MDMA-based network, while the highest outage probability is observed in the NOMA-based network. This is because in NOMA, in a power-constrained  system, both X and Y can only be transmitted simultaneously only using the maximum transmission power of the node, which greatly increases the difficulty of successful decoding.

\begin{figure}[h!]
	\centerline{\includegraphics[width=3.5in]{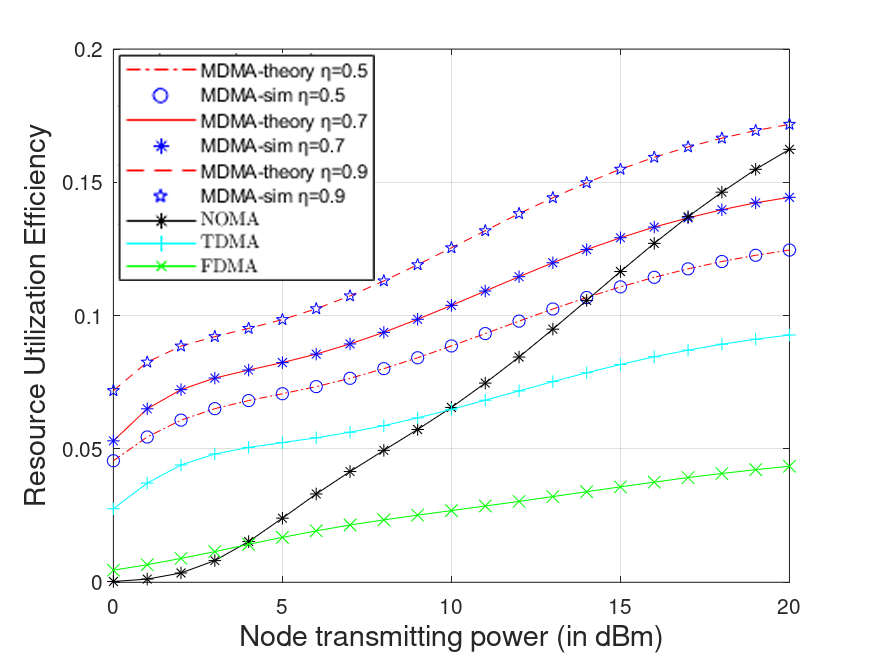}}
	\caption{Resource utilization efficiency of system vs. node transmitting power.}
	\label{fig3}
\end{figure}

According to Fig.\ref{fig3}, it can be observed that the MDMA system achieves the highest resource utilization when the transmission power of nodes $P_{T}$ is less than 14dBm. When $P_{T}$ exceeds 14dBm, the NOMA system gradually surpasses the MDMA systems with $\eta=0.5$ and $\eta=0.7$, but it still remains lower than the MDMA system with $\eta=0.9$. Additionally, it can be seen that the FDMA system consistently maintains a lower resource utilization. This is due to the fact that FDMA utilizes two sets of power and two sets of bandwidth, while the remaining MA systems are assumed to use just one set of resources. In the NOMA system, the resource utilization increases rapidly with the increase of $P_{T}$. This is because the NOMA system can transmit both X and Y simultaneously using one set of resources. This significantly improves the transmission efficiency, especially at high signal-to-noise ratios, and fully utilizes the available resources.

Moreover, it can be observed from all the result figures that the theoretical results and the simulation results are well-matched.
\section{Conclusion}
This paper combines cooperative communication MRC network with MDMA. By applying the MGF and STM, the theoretical algorithm-based expressions for the outage probability, resource utilization efficiency, and the average number of time slots required for each image transmission are derived. Furthermore, through numerous simulation results, it is demonstrated that our work outperforms traditional approaches.

\end{document}